\documentstyle[12pt]{article}
\begin{document}

\author{Amand Faessler$^1$, A. J. Buchmann$^1$, M. I. Krivoruchenko$^{1,2}$ and\\
B. V. Martemyanov$^2$ \\
{\small $^1${\it Institut f\"ur Theoretische Physik, Universit\"at
T\"ubingen, Auf der Morgenstelle 14 }}\\
{\small {\it D-72076 T\"ubingen, Germany}}\\
{\small $^2${\it Institute for Theoretical and Experimental Physics,
B.Cheremushkinskaya 25}}\\
{\small {\it 117259 Moscow, Russia}}}
\title{Nuclear Matter with a Bose Condensate of Dibaryons in Relativistic
Mean-Field Theory}
\date{}
\maketitle

\begin{abstract}
If sufficiently light dibaryon resonances exist, a Bose condensate of
dibaryons can occur in nuclear matter before the quark-hadron phase
transition. Within a relativistic mean-field model we show that heterophase
nuclear-dibaryon matter is for a wide set of parameters energetically more
favorable than normal nuclear matter. Production of dibaryons is, however,
relatively suppressed as compared to estimates based on the model of
non-interacting nucleons and dibaryons.
\end{abstract}

\vspace{1 cm}

The possibility of existence of dibaryon resonances was investigated in the
last two decades experimentally and theoretically. The most promising
candidates for experimental searches are those dibaryons which have a small
width. In 1977 R.Jaffe predicted \cite{Jaf} the existence of a loosely bound
dihyperon, $H$, with a mass just below the threshold for two-lambda decay.
Calculations of the $H$-particle mass within $QCD$-inspired models [2-4]
showed that the existence of a dihyperon near the $\Lambda \Lambda $%
-threshold is plausible. Dibaryons with exotic quantum numbers, which have a
small width due to zero coupling to the $NN$-channel, are of special
interest [5-7]. A method for searching narrow, exotic dibaryon resonances in
the double proton-proton bremsstrahlung reaction is discussed in Ref. \cite
{Ger}. Data from pion double charge exchange (DCE) reaction on nuclei \cite
{Bil, HCl} exhibit a peculiar energy dependence at the total pion energy of $%
190$ $MeV$ that can be interpreted \cite{Mar} as evidence for the existence
of a narrow $d^{\prime }$ dibaryon with quantum numbers $T=0$, $J^P=0^{-}$.
Recent experiments at TRIUMPF (Vancouver) and CELSIUS (Uppsala) seem to
support the existence of the $d^{\prime }$ dibaryon \cite{Bro}.

The properties of nuclear matter with admixture of multiquark clusters are
discussed in Ref. \cite{Bal}. A dibaryon Bose condensate in interiors of
neutron stars decreases the maximum masses of neutron stars \cite{Kri}. In a
recent paper \cite{Buc} an exactly solvable model for a one-dimensional
system of fermions interacting through a potential, which leads to a
resonance in the two-fermion channel, is constructed. The behavior of this
system can be interpreted in terms of a Bose condensation of the two-fermion
resonances.

There is no dibaryon condensate in ordinary nuclei. From this one can
conclude that the masses of dibaryons coupled to the $NN$-channel should be
greater than 
\begin{equation}
\label{f.1}m_D>2\mu _N=2(m_N+\varepsilon _F)=1.96{\cal \;\;}GeV 
\end{equation}
where $\mu _N$ is the chemical potential of nucleons and $\varepsilon _F=40$ 
$MeV$ is the Fermi energy of nucleons in nuclei. Here, the ideal gas
approximation for nucleons and dibaryons and the assumption that the shell
model potential for dibaryons is twice as deep as the one for nucleons have
been used.

The $d^{\prime }$ dibaryon is coupled to the $NN\pi $ channel only. In the
nuclear medium, the reaction $nd^{\prime }\leftrightarrow nnp$ is possible.
In nuclei the equilibrium condition for the chemical potentials has the form 
$\mu _n+m_D=2\mu _n+\mu _p $. Since $\mu _p\approx \mu _n$, we arrive at the
same inequality (\ref{f.1}).

A Bose condensate of dibaryons can presumably be formed at high densities
when relativistic effects for nucleons become important. In order to
describe such a system, we should go beyond non-relativistic many-body
theory. The relativistic field-theoretical Walecka model \cite{Wal} is known
to be very successful in describing properties of infinite nuclear matter
and of ordinary nuclei throughout the periodic table. In this paper we study
the influence of narrow dibaryon resonances on nuclear matter in the
framework of the Walecka model in the mean-field approximation.

The Lagrangian of the model contains nucleons interacting through $\omega $-
and $\sigma $-meson exchanges. We add to the Lagrangian dibaryons
interacting with nucleons and each other through $\omega $- and $\sigma $%
-meson exchanges also. Inclusion of dibaryons entails uncertainties
connected to the lack of reliable information on dibaryon masses and
coupling constants. However, many conclusions can be drawn on quite general
grounds without knowing precise values for the newly added parameters. The
Lagrangian density is given by 
\begin{equation}
\label{f.2}
\begin{array}{c}
{\cal L}=\bar \Psi (i\partial _\mu \gamma _\mu -m_N-g_\sigma \sigma
-g_\omega \omega _\mu \gamma _\mu )\Psi +\frac 12(\partial _\mu \sigma
)^2-\frac 12m_\sigma ^2\sigma ^2 \\ -\frac 14F_{\mu \nu }^2+\frac 12m_\omega
^2\omega _\mu ^2+(\partial _\mu -ih_\omega \omega _\mu )\varphi
^{*}(\partial _\mu +ih_\omega \omega _\mu )\varphi -(m_D+h_\sigma \sigma
)^2\varphi ^{*}\varphi +{\cal L}_c. 
\end{array}
\end{equation}
Here, $\Psi $ is the nucleon field, $\omega _\mu $ and $\sigma $ are fields
of the $\omega $- and $\sigma $-mesons, $F_{\mu \nu }=\partial _\nu \omega
_\mu -\partial _\mu \omega _\nu $, $\varphi $ is the dibaryon
isoscalar-scalar (or isoscalar-pseudoscalar) field. The values $m_\omega \ $
and $m_\sigma $ are the $\omega $- and $\sigma $-meson masses and the values 
$g_\omega $, $g_\sigma $, $h_\omega $, $h_\sigma $ are coupling constants of
the $\omega $- and $\sigma $-mesons with nucleons ($g$) and dibaryons ($h$).

The term ${\cal L}_c$ describes conversion of dibaryons into nucleons. The $%
H $-particle is coupled to the $NN$-channel through a double weak decay and $%
{\cal L}_c=O(G_F^2).$ For the nonstrange $d_1$ dibaryon \cite{Khr} and the $%
d^{\prime }$ dibaryon, we neglect possible virtual transitions {\it e.g.} to
the $NN\sigma $ channel. The on-shell couplings for these dibaryons are
small. The exotic $d_1$ dibaryon decays to the $NN\gamma $-channel only, and
so ${\cal L}_c=O(\alpha )$. The $d^{\prime }$ dibaryon decays to the $NN\pi $
channel. Due to Adler's consistency condition \cite{Adl} ${\cal L}_c$ $%
\propto \partial _\mu {\bf \pi }$ . In the mean-field approximation $%
\partial _\mu {\bf \pi }=0$, and the term ${\cal L}_c$ does not modify the
mean-field equations. In what follows we set ${\cal L}_c=0$. The effect of a
small term ${\cal L}_c$ reduces to providing a chemical equilibrium with
respect to transitions between dibaryons and nucleons \cite{Shu}.

The field operators can be expanded in $c$-numbers and operator parts: $%
\omega _\mu =g_{\mu 0}\omega _c+\hat \omega _\mu ,$ $\sigma =\sigma _c+\hat
\sigma ,$ $\varphi =\varphi _c+\hat \varphi ,$ and $\varphi ^{*}=\varphi
_c^{*}+\hat \varphi ^{*}$. The $\sigma $-meson mean field determines the
effective nucleon and dibaryon masses in the medium: $m_N^{*}=m_N+g_\sigma
\sigma _c$ and $m_D^{*}=m_D+h_\sigma \sigma _c.$ The baryon number current
has the form $j_\mu ^B=j_\mu ^N+2j_\mu ^D\ $ where $j_\mu ^N=\bar \Psi
\gamma _\mu \Psi $ and $j_\mu ^D=\varphi ^{*}i\stackrel{\leftrightarrow }{%
\partial } _\mu \varphi -2h_\omega \omega _\mu \varphi ^{*}\varphi .$ The $%
\omega $-mesons are coupled to the current $j_\mu ^\omega =g_\omega
j_\mu^N+h_\omega j_\mu ^D$.

The nucleon vector and scalar densities are defined by the expectation
values $\rho _{NV}=<\bar \Psi (0)\gamma _0\Psi (0)>$ and $\rho _{NS}=<\bar
\Psi (0)\Psi (0)>$. The scalar density of the dibaryon condensate is given
by $\rho _{DS}^c=\left| <\varphi (0)>\right| ^2$. The time evolution of the
dibaryon condensate $\varphi $-field is determined by the dibaryon chemical
potential $\mu _D$ 
\begin{equation}
\label{f.3}\varphi _c(t)=e^{-i\mu _Dt}\sqrt{\rho _{DS}^c}. 
\end{equation}
The vector density of the dibaryon condensate is given by $\rho _{DV}^c=2\mu
_D^{*}\rho _{DS}^c$ where $\mu _D=\mu _D^{*}+h_\omega \omega _c.$

The existence of a dibaryon condensate depends on the values of the coupling
constants of dibaryons with the $\omega $- and $\sigma $ -mesons. The $%
\omega $- and $\sigma $- meson coupling constants $h_\omega $ and $h_\sigma $
enter the dibaryon-dibaryon Yukawa potential 
\begin{equation}
\label{f.4}V(r)=\frac{h_\omega ^2}{4\pi }\frac{e^{-m_\omega r}}r-\frac{%
h_\sigma ^2}{4\pi }\frac{e^{-m_\sigma r}}r. 
\end{equation}
The interaction energy for dibaryons in the condensate is for a constant
density distribution $\rho _D({\bf x})=$ $\rho _D$ equal to 
\begin{equation}
\label{f.5}W=\frac 12\int d{\bf x}_1d{\bf x}_2\rho _D({\bf x}_1)\rho _D({\bf %
x}_2)V(|{\bf x}_1-{\bf x}_2|)=2\pi N_D\rho _D(\frac{h_\omega ^2}{4\pi
m_{_\omega }^2}-\frac{h_\sigma ^2}{4\pi m_{_\sigma }^2}) 
\end{equation}
where $N_D$ is the total number of dibaryons. A negative $W$ would imply
instability of the system against compression. The value $W$ is positive and
the system is stable for 
\begin{equation}
\label{f.6}\frac{h_\omega ^2}{4\pi m_{_\omega }^2}>\frac{h_\sigma ^2}{4\pi
m_{_\sigma }^2}. 
\end{equation}
In a nonrelativistic theory for systems of interacting bosons \cite{Abr} and
in the model considered, the requirement of stability is equivalent to the
requirement of a positive value for the square of the sound velocity ($%
a_s^2>0$).

The $H$-particle interactions are studied on the basis of the
non-relativistic quark cluster model [3, 22-24] which is successful in
describing the $NN$-phase shifts. The calculation of the interaction
integral (\ref{f.5}) with the adiabatic $HH$-potential \cite{Sak} gives a
negative energy, so the $H$-dibaryon condensate is probably unstable against
compression. The coupling constants of the mesons with the $H$-particle can
be fixed by fitting the depth and the position for minimum of the $HH$
-potential to give $h_\omega ^2=603.7$ and $h_\sigma ^2=279.2$.

The coupling constants of the mesons with the $d_1$ and $d^{^{\prime }}$
dibaryons are unknown. The $\omega $- and $\sigma $-mesons interact with
nucleons and pions inside the dibaryon. For dibaryons decaying into the $NN$%
-channel, the $\sigma-D $ and $\omega-D $ couplings are in nonrelativistic
approximation two times greater than for nucleons: $h_\omega =2g_\omega $
and $h_\sigma =2g_\sigma $. The scalar charge is, however, suppressed by the
Lorentz factor. For standard parameters of the Walecka model$\ $\cite{Wal}, $%
m_\sigma =520$ $MeV$, $g_\omega ^2=190.4$, and $g_\sigma ^2=109.6$, the
inequality (6) becomes $98.85(\frac{h_\omega }{ 2g_\omega })^2GeV^{-2}>129.0(%
\frac{h_\sigma }{2g_\sigma })^2GeV^{-2}.$ With these additive estimates, the
inequality (6) is not fulfilled. The exchange current contributions to the
meson couplings with dibaryons, which violate additivity, are analysed in
Ref. \cite{Ama}. At present no definite conclusions concerning the stability
of the $d_1$ and $d^{^{\prime }}$ dibaryon matter can be drawn.

The mean-field solutions of the equations of motion corresponding to the
Lagrangian density (2) are obtained by neglecting the operator parts of the
meson fields. For the $\omega $- and $\sigma $-meson mean fields, we get the
following expressions 
\begin{equation}
\label{f.7}\omega _c=\frac{g_\omega \rho _{NV}+h_\omega 2\mu _D^{*}\rho
_{DS}^c}{m_\omega ^2}, 
\end{equation}
\begin{equation}
\label{f.8}\sigma _c=-\frac{g_\sigma \rho _{NS}+h_\sigma 2m_D^{*}\rho
_{DS}^c }{m_\sigma ^2}. 
\end{equation}

Substituting expression (\ref{f.3}) into the equation of motion for the
dibaryon field $\varphi $, we get $\mu _D^{*}=m_D^{*}.$ The nucleon and
dibaryon chemical potentials have the form $\mu _N=E_F^{*}+g_\omega \omega
_c $ and $\mu _D=m_D^{*}+h_\omega \omega _c,$ where $E_F^{*}=\sqrt{m_N^{*2}+%
{\bf k}_F^2 }$ is the Fermi energy of nucleons with the effective mass $%
m_N^{*}$.

The self-consistency condition for the effective nucleon mass can be
transformed to a form equivalent to that in the standard Walecka model: 
\begin{equation}
\label{f.9}m_N^{*}=\tilde m_N-\frac{g_\sigma ^2}{m_\sigma ^2}\rho _{NS}, 
\end{equation}
where $\tilde m_N=m_N(\rho _{DV}^{c,\max }-\rho _{DV}^c)/\rho _{DV}^{c,\max
} $ and $\rho _{DV}^{c,\max }=m_Nm_\sigma ^2/(g_\sigma h_\sigma )=0.1507( 
\frac{2g_\sigma }{h_\sigma })fm^{-3}.$ If the densities $\rho _{TV}$ and $%
\rho _{DV}^c$ are fixed, equation (\ref{f.9}) allows to find the effective
nucleon mass $m_N^{*}$. Solutions to Eq.(\ref{f.9}) exist for arbitrary
total density $\rho _{TV}$ and dibaryon density $\rho _{DV}^c<\rho
_{DV}^{c,\max }$when the value $\tilde m_N$ is positive.

It is clear that the fraction of dibaryons should increase when the
difference $2\mu _N-\mu _D$ is positive and $\rho _{DV}^c=0$. If the
difference $2\mu _N-\mu _D$ is negative and the system consists of dibaryons
only, production of nucleons is energetically favorable. If the difference $%
2\mu _N-\mu _D=0\;$and increases with the dibaryon fraction, small
fluctuations take the system away from equilibrium. These types of states
are unstable.

There are three stable cases: (i) The homophase nuclear matter: $2\mu _N-\mu
_D<0$ and $\rho _{DV}^c=0;\;$(ii) the homophase dibaryon matter: $2\mu
_N-\mu _D>0$ and $2\rho _{DV}^c=\rho _{TV};$\ (iii)\ the heterophase
nuclear-dibaryon matter: $2\mu _N-\mu _D=0$ and $\frac{d(2\mu _N-\mu _D)}{
d\rho _{DV}^c}\left| _{\rho _{TV}}\right. <0.$ Small fluctuations around
these states lead the system back to the equilibrium points.

In Fig.1 we show the critical density for occurrence of a Bose condensate of
non-strange dibaryons for $h_\omega =2g_\omega $ as a function of the $%
\sigma $-meson coupling constant with dibaryons. The critical density is
determined from the equation $2\mu _N-\mu _D=0.$

In Fig.2 (a) we show the nucleon effective mass $m_N^{*}$ versus the
dibaryon fraction $2\rho _{DV}^c/\rho _{TV}$ for $h_\omega =2g_\omega $ and $%
h_\sigma =1.6g_\sigma $. The behavior of $m_N^{*}$ as a function of the
dibaryon fraction does not depend on the $m_D$, since the dibaryon mass does
not enter the self-consistency condition (\ref{f.9}) directly. In Fig.2 (b)
we show the difference for the chemical potentials versus the dibaryon
fraction.

Mean-field solutions exist at all densities $\rho _{TV}$ for sufficiently
small densities of dibaryons, $\rho _{DV}^c$$<\rho _{DV}^{c,\max }$. This
means that we can always investigate the stability of normal nuclear matter
with respect to dibaryon condensate formation. Small total baryon number
densities correspond to a stable equilibrium of type (i), at higher
densities a stable equilibrium (iii) occurs. When the density $\rho _{TV}$
is high, dibaryon production is energetically favorable. The mean-field
solutions disappear, however, before the system reaches an equilibrium.

In Fig. 3 we show the energy per nucleon and the pressure versus the total
baryon number density for some possible dibaryons. The effect of zero
compressibility for heterophase nuclear-dibaryon matter present in the ideal
gas approximation reveals itself through the softening of the equation of
state (EOS). Notice that the pressure of the heterophase system obeys the
basic inequality \cite{Kad}\ $\partial p/\partial \rho _{TV}\geq 0\;$ of
statistical mechanics. One can verify that in the model considered, the
hydrostatic pressure coincides with the thermodynamic pressure. The $H$%
-particles are formed at a lower density, since the $\omega -H$ coupling
constant $h_\omega /(2g_\omega )=0.89$ is relatively small. The energy of
the $H$-particles in the positive $\omega $-meson mean field is lower, so
the production of the $H $-particles is energetically more favourable.

The qualitative estimates based on a model for non-interacting nucleons and
dibaryons show that in normal nuclear matter a dibaryon Bose condensate does
not exist provided the inequality (\ref{f.1}) is satisfied. A more accurate
estimate can be made on the basis of the relativistic mean-field model (\ref
{f.2}). From the requirement of absence of a dibaryon Bose condensate for $%
\rho _{TV}\leq \rho _0=0.15\;fm^{-3}$, we get for $h_\omega =2g_\omega $ a
constraint 
\begin{equation}
\label{f.10}m_D>1.89\;GeV. 
\end{equation}
This constraint is valid provided that the dibaryon matter is stable against
compression. In such a case, the $d_1$ resonance \cite{Khr} with a mass $%
m_D=1.92$ $GeV$ does not affect properties of ordinary nuclei.

Phase transitions of nuclear matter to strange quark matter \cite{Wit, Far}
have been widely discussed in the literature (for a recent review see \cite
{Las}). Dense nuclear matter with a dibaryon Bose condensate can exist as an
intermediate state below the quark-gluon phase transition. This is the case
when dibaryon matter is stable against compression. If dibaryon matter is
unstable, the creation of dibaryons can be a possible mechanism for the
phase transition to quark matter. The energetically favourable compression
of $H$-matter can lead to the formation of strange matter. It would be
interesting to check astrophysical data for the presence of a dibaryon
condensate in the interiors of massive neutron stars as well as possible
signatures of their instability caused by dibaryons.

\vspace{1cm}

The authors are grateful to S. Gerasimov, M. Kirchbach and M. G. Schepkin
for useful discussions. Two of us (B. V. M. and M. I. K.) are grateful to
RFBR for Grant No. 94-02-03068 and Neveu-INTAS for Grant No. 93-0023. M. I.
K. acknowledges hospitality of Institute for Theoretical Physics of
University of Tuebingen, Alexander von Humboldt-Stiftung for a
Forschungsstipendium, and INTAS for Grant No. 93-0079.

\begin{center}
\newpage\ 

\ {\bf FIGURE CAPTIONS}
\end{center}

{\bf Fig.1.} The critical density for occurrence of a Bose condensate of
dibaryons versus the $\sigma $-meson coupling constant $h_\sigma $ for $%
m_D=1.88$ $GeV$ ($=2m_N$; the long dashed curve No. $1$), $1.96$ $GeV$ (the
solid curve No. $2$), {\it etc.} with a step $80$ $MeV$. The results for
dibaryons $d_1(1920)$ and $d^{^{\prime }}(2060)$ are also shown (the dashed
curves). The dibaryon matter is stable against compression when the square
of the sound velocity $a_s$ is positive. This is the case for $h_\sigma
/(2g_\sigma )<0.8754$. The value $\rho _0=0.15\;fm^{-3}$ is the saturation
density for nuclear matter. For $2m_N\leq m_D\leq 1.89\;GeV$, we start at
zero density from a heterophase nuclear-dibaryon matter. With increasing the
density, the matter can be transformed to a homophase nuclear matter and
then again to a heterophase nuclear-dibaryon matter. For $m_D>1.89\;GeV$, we
start at zero density from a homophase nuclear matter which converts with
increasing the density (at $\rho _{TV}>\rho _0$ for $h_\sigma /(2g_\sigma
)<0.8754$) to a heterophase nuclear-dibaryon matter. The occurrence of the $%
H $-particles is denoted by the cross.

{\bf Fig.2}. (a) The effective nucleon mass $m_N^{*}$ in $GeV$ versus the
dibaryon fraction $2\rho _{DV}^c/\rho _{TV}\ $in the heterophase
nuclear-dibaryon matter. The results do not depend on the dibaryon mass. (b)
The difference $2\mu _N-\mu _D$ between the two nucleon chemical potentials
and the dibaryon chemical potential versus the dibaryon fraction $2\rho
_{DV}^c/\rho _{TV}$. The results are given for the total baryon densities $%
1, $ $2,$ $3,$ $4,$ $5,$ and $6$ times greater than the saturation density.
The normal nuclear matter is stable when $2\mu _N-\mu _D<0\ $and $\rho
_{DV}^c=0.\;$An intersection of a curve with a negative slope with the
horizontal line $2\mu _N-\mu _D=0$ indicates occurrence of a stable
equilibrium in the heterophase nuclear-dibaryon matter. The results are
given for $m_D=1.96$\ $GeV.$ The dibaryon mass does not enter the
self-consistency condition (9) and enters linearly in the difference $2\mu
_N-\mu _D$, so the curves for other dibaryon masses can be obtained by
vertical parallel displacements.$\ $The results for $m_D=2.06$ $GeV$ ($%
d^{\prime }$-dibaryon) can be obtained {\it e.g. }by a $100$ $MeV$ negative
shift, {\it etc.}

{\bf Fig.3}. The pressure (left scale) and the energy per baryon (right
scale) versus the total baryon number density $\rho _{TV}=\rho _{NV}+2\rho
_{DV}^c$ for normal nuclear matter (solid lines) and for the heterophase
nuclear-dibaryon matter (dashed lines) for the $d_1(1920)$ and $d^{^{\prime
}}(2060)$ dibaryons at $h_\omega =2g_\omega $ and $h_\sigma /(2g_\sigma
)=0.8 $. The dibaryon condensates occur at $\rho _{TV}/\rho _0=2.05$ and$%
\;3.15\ $for the $d_1$ and $d^{^{\prime }}.$ The dibaryon Bose condensation
softens the EOS for nuclear matter. The $H$-dibaryons are formed at $\rho
_{TV}/\rho _0=2.74$ when the meson coupling constants are fixed by fitting
the adiabatic $HH$-potential \cite{Sak}.\ In this case the $H$-matter is
unstable against compression, providing a transition to strange matter.

\end{document}